# Title: Large Impacts around a Solar Analog Star in the Era of Terrestrial Planet Formation

**Authors:** Huan Y. A. Meng[1], Kate Y. L. Su[2], George H. Rieke[1,2], David J. Stevenson[3], Peter Plavchan[4], Wiphu Rujopakarn[2,5,6], Carey M. Lisse[7], Saran Poshyachinda[8], Daniel E. Reichart[9]

**Affiliations:**

[1] Lunar and Planetary Laboratory and Department of Planetary Sciences, University of Arizona, 1629 E University Blvd, Tucson, AZ 85721.

[2] Steward Observatory and Department of Astronomy, University of Arizona, 933 N Cherry Ave, Tucson, AZ 85721.

[3] Division of Geological and Planetary Sciences, California Institute of Technology, MC 170-25, 1200 E California Blvd, Pasadena, CA 91125.

[4] NASA Exoplanet Science Institute, California Institute of Technology, MC 100-22, 770 S Wilson Ave, Pasadena, CA 91125.

[5] Department of Physics, Faculty of Science, Chulalongkorn University, 254 Phayathai Rd, Pathumwan, Bangkok 10330, Thailand.

[6] Kavli Institute for the Physics and Mathematics of the Universe (WPI), Todai Institute for Advanced Study, University of Tokyo, 5-1-5 Kashiwanoha, Kashiwa, 277-8583, Japan.

[7] Space Department, Applied Physics Laboratory, Johns Hopkins University, 11100 Johns Hopkins Rd, Laurel, MD 20723.

[8] National Astronomical Research Institute of Thailand (Public Organization), Ministry of Science and Technology, 191 Siriphanich Bldg, Huay Kaew Rd, Muang District, Chiang Mai 50200, Thailand.

[9] Department of Physics and Astronomy, Campus Box 3255, University of North Carolina at Chapel Hill, Chapel Hill, NC 27599.

**Abstract**: The final assembly of terrestrial planets occurs via massive collisions, which can launch copious clouds of dust that are warmed by the star and glow in the infrared. We report the real-time detection of a debris-producing impact in the terrestrial planet zone around a 35-million year-old solar analog star. We observed a substantial brightening of the debris disk at 3-5 μm, followed by a decay over a year, with quasi-periodic modulations of the disk flux. The behavior is consistent with the occurrence of a violent impact that produced vapor out of which a thick cloud of silicate spherules condensed that were ground into dust by collisions. These results demonstrate how the time domain can become a new dimension for the study of terrestrial planet formation.

**One Sentence Summary:** We observe production and time evolution of dust clouds around a solar analog star, resulting from a violent collision of large bodies in the era of terrestrial planet formation.

**Main Text:** Circumstellar disks are where planetary systems form and evolve. Gas rich and optically thick protoplanetary disks are born together with young stars but dissipate within a few million years (Myr), setting the timescale for gas giant planet formation (*1*). Dusty, optically thin debris disks then emerge, sustained by fragmentation of colliding planetesimals (*2*). They are warmed by their stars and light up in the infrared, allowing them to be detected over the entire lifetimes of main sequence stars. Consequently, debris disks are ideal tools to search for phases occurring in other planetary systems that are analogous to major events in the evolution of the solar system, such as the formation of terrestrial planets. Dynamical simulations and meteoritics indicate that the end stage of terrestrial planet formation, from ~30 to ~100 Myr (e.g., *3*, *4*), is marked by frequent large impacts, up to the scale of the one leading to the formation of the Moon (*5-7*). Such events can produce a huge amount of dust, which may dramatically increase the infrared emission from the debris disk (e.g., *8*).

ID8 (2MASS J08090250-4858172) is a young solar analog star with spectral type G6V and solar metallicity in the 35-Myr-old open cluster NGC 2547 (*9*). It emits strongly in the infrared, with a fractional disk luminosity of $L_{disk}/L_*=3.2\times10^{-2}$ (*10*). The mid-infrared spectrum of ID8 (observed in 2007) shows strong crystalline silicate features from 8 to 30 µm, indicative of very fine dust particles (Fig. 1); models for the observed spectral energy distribution (SED) require sub-µm-sized amorphous silicate dust as well (cf. *10*). Particles are blown out by radiation pressure and lost to the system if the ratio of radiation to gravitational forces is larger than 0.5. For ID8, the critical radius (*11*) below which non-porous silicate grains are lost is ~0.5 µm. The tiniest particles in the disk are smaller than this limit, suggesting ongoing dust replenishment in the system.

The disk emission was recently found to vary on a yearly timescale (*8*). Collisional cascades among planetesimals sustain most debris disks. However variations this rapid cannot be supported by this means because the cascade timescales for significant variations in dust production are at least a few hundred orbits (*8*). To explore the origin of the variations, we have used the Infrared Array Camera (IRAC, *12*) onboard the Spitzer Space Telescope to monitor the ID8 system. The observations extended from May 25, 2012 to August 23, 2013, providing a total time baseline of 454 days with a 157-day gap between visibility windows. At the same time, intensive optical monitoring of ID8 was obtained from the ground in the V and Cousins I ($I_C$ hereafter) bands where we found that the output of the star is stable within ~1.5% RMS (supplementary online text).

Because the stellar contribution is effectively constant in time, we fitted its spectrum and removed it from the total (star + disk) infrared fluxes to obtain the light curves of the debris disk, as shown in Fig. 2. This revealed an average disk color of [3.6]-[4.5]=1.00, corresponding to a blackbody temperature of 730 K, consistent with the temperatures found in previous analyses of the infrared spectroscopy (*10*). We calculated the expected color trend of the entire ID8 system for possible causes of the disk variations and compared with our data. The result is not conclusive within the errors, but suggests that a combination of changing dust temperature and dust emitting area, or in area alone rather than purely in temperature, may be responsible for the disk variations (supplementary online text).

In the following analysis, we focus on the 4.5 µm data where we have observations from both years and the disk is measured at a higher signal-to-noise ratio. The data at 3.6 µm show consistent behavior, although at lower signal-to-noise. In 2012, the disk flux density stayed near 2.2 milli-Jansky (mJy) despite ~10% variations. However, at the start of 2013 it had brightened to above 3.0 mJy, indicating a significant increase of the amount of dust, probably from a new impact before 2013. This elevated level then decayed throughout 2013. An exponential fit suggests a decay timescale of ~370 days at both wavelengths. This is too fast to be reconciled with decades-long collisional cascades (*8*), and is too slow for the direct radiative blowout of tiny particles, which should take <30 days at the orbit derived below for the disk.

To better understand the impact, we estimate the disk mass based on a fit to the entire mid-infrared spectrum, which is dominated by small grains. Since the new debris in the disk has not reached equilibrium in a full collisional cascade, we could not make the conventional assumption of a power law size distribution. For emitting grains of 0.5 μm in radius we found a disk mass of $1.1 \times 10^{19}$ kg, which is a lower limit since it ignores larger particles. An independent estimate assuming a power law grain size distribution up to 1 mm obtains an identical mass estimate (*10*). This mass, if the grains were compacted into a solid body, is equivalent to a ~180-km diameter asteroid (of density 3700 kg m$^{-3}$). Given the estimated mass and particle size, the ID8 disk may be optically thick, in which significant mass could be obscured and unseen.

Considering the decay in 2013, and assuming it applies to the full spectrum, we estimated the mass loss rate to be at least $10^{11}$ kg s$^{-1}$. The spectrum was obtained in 2007; however, the Spitzer and WISE photometric points (Fig. 1) imply that the mid-infrared spectrum had a similar shape at least from January 2004 through November 2010. That is, it appears that the small grains, with a net volume of a ~180-km diameter asteroid, were lost from the system and would have to be replenished on a decadal (or shorter) timescale to maintain the mid-infrared spectrum. Though arising from different physical processes, the mass loss rate is 5-7 orders of magnitude greater than the dust mass loss rates of comets Hale-Bopp (*13*) and Halley (*14*), among the dustiest comets known in the solar system, and is 3 orders of magnitude greater than that of the evaporating planet KIC 12557548b (*15*).

The variability of the infrared emission of ID8 is much too fast to arise in a conventional debris disk sustained by a collisional cascade (*8*). We hypothesize that the impact responsible for the increased disk emission in 2013 involved two large bodies and was sufficiently violent to yield a silica-rich vapor plume. Glassy silicate spherules will condense from the vapor with diverse forms (e.g., *16*), consistent with the presence of amorphous silicates in the SED model (*10*). In this case, there may have been temporal spectral features after the new impact, but would have been missed because we do not have new mid-infrared spectrum in 2013. The condensation process has been modeled in (*17*), which shows that the typical spherule size depends sensitively on the circumstances of the impact, particularly its velocity, but ranges from about 10 μm to 1 mm. The condensates are produced quickly over several hours. Initially, the cloud of spherules will not radiate efficiently in the mid-infrared because the total surface area of all the spherules is small. However, they will break each other down through collisions, which generate the observed μm-sized or smaller particles as daughter products. The infrared output will rise as the mass is distributed into many small grains, making the consequences of the initial impact visible as an increase in the infrared emission.

Because the size distribution of condensate spherules is strongly peaked around the average (*17*), we treat them as being equal in size. Then, a rough estimate of the timescale for destroying them (and hence for the decay of the debris cloud) can be obtained by attributing all the disk mass to them immediately after the impact and assuming that they are removed by breakdown in a collisional cascade with eventual ejection, when their daughter particles are sub-μm in size, by radiation pressure. We found that different models for the destruction rate of such spherules give consistent timescales (*18*, *19*). To order of magnitude, the range of decay timescales as a function of spherule size from 10 to 1000 μm is 100 days to 10 years. The observed decay time of the ID8 disk corresponds to a spherule size of ~100 μm, which corresponds to an impact velocity of 15-

18 km s$^{-1}$ (for a 100-1000 km diameter body impacting on an even larger one, *17*). Thus, the one-year exponential decay timescale is a natural result for a system where seed grains have condensed from a vapor cloud and are destroying themselves through collisions. An analysis with some similarities to ours for the bright debris disk of HD 172555 (*20*) found that dust created in a hypervelocity impact will have a size slope of ~-4, in agreement with the fits of (*10*) to the infrared spectrum of ID8.

After the exponential decay is removed from the data ("detrending"), the light curves at both wavelengths appear to be quasi-periodic. The regular recovery of the disk flux and lack of extraordinary stellar activity essentially eliminate coronal mass ejection (*21*) as a possible driver for the disk variability. We employed the SigSpec algorithm (*22*) to search for complex patterns in the detrended, post-impact 2013 light curve. The analysis identified two significant frequencies with comparable amplitudes, whose periods are $P_1$=25.4±1.1 days and $P_2$=34.0±1.5 days (Fig. 3(A)) and are sufficient to reproduce qualitatively most of the observed light curve features (Fig. 3(B)). The quoted uncertainties (*23*) do not account for systematic effects due to the detrending, and thus are lower limits to the real errors. Other peaks with longer periods in the periodogram are aliases, or possibly reflect long-term deviation from the exponential decay. These artifacts make it difficult to determine if there are weak real signals near those frequencies.

We now describe the most plausible interpretation of this light curve that we have found. The two identified periods have a peak-to-peak amplitude of ~6×10$^{-3}$ in fractional luminosity, which provides a critical constraint for models of the ID8 disk. In terms of sky coverage at the disk distance inferred from the infrared SED, such an amplitude requires disappearance and reappearance every ~30 days of the equivalent of an opaque, stellar facing "dust panel" of radius ~110 Jupiter radii. One possibility is that the disk flux periodicity arises from recurring geometry that changes the amount of dust that we can see. At the time of the impact, fragments get a range of kick velocities when escaping into interplanetary space. This will cause Keplerian shear of the cloud (*24*), leading to an expanding debris concentration along the original orbit (supplementary online text). If the ID8 planetary system is roughly edge-on, the longest dimension of the concentration will be parallel to our line of sight at the greatest elongations, and orthogonal to the line of sight near conjunctions to the star. This would cause the optical depth of the debris to vary within an orbital period, in a range on the order of 1-10 according to the estimated disk mass and particle sizes. Our numerical simulations of such dust concentrations on moderately eccentric orbits are able to produce periodic light curves with strong overtones. $P_2$ and $P_1$ should have a 3:2 ratio if they are the first and second order overtones of a fundamental, which is consistent with the measurements within the expected larger errors (<2σ or better). In this case, the genuine period should be 70.8±5.2 days (lower limit errors), a value where it may have been submerged in the periodogram artifacts. This period corresponds to a semi-major axis of ~0.33 AU, which is consistent with the temperature and distance suggested by the spectral models (*10*).

Despite the peculiarities of ID8, it is not a unique system. In 2012 and 2013 we monitored four other "extreme debris disks" (with disk fractional luminosity ≥10$^{-2}$) around solar-like stars with ages of 10-120 Myr. Various degrees of infrared variations were detected in all of them. The specific characteristics of ID8 in the time domain, including the yearly exponential decay, additional more rapid weekly to monthly changes, and color variations, are also seen in other systems. This opens up the time domain as a new dimension for the studies of terrestrial planet formation and collisions outside the solar system. The variability of many "extreme debris disks" in the era of the final buildup of terrestrial planets may provide new possibilities to understand the early solar system and habitable planet formation (e.g., *25*).

**Acknowledgments:** H.M, K.S, and G.R would like to thank Renu Malhotra and András Gáspár for valuable discussions. This work is based on observations made with the Spitzer Space

Telescope, which is operated by the Jet Propulsion Laboratory, California Institute of Technology under a contract with NASA. Support for this work was provided by NASA through an award issued by JPL/Caltech, and by NASA grant NNX13AE74G. All data are publicly available through NASA/IPAC Infrared Science Archive.

**Supplementary Materials:**

Optical Observations of ID8

Spitzer Observations and Data Analysis

Frequency Spectrum Analysis

Frequency Spectrum Interpretation

Scale of the Impact

Figure S1-S4

References (*28-45*)

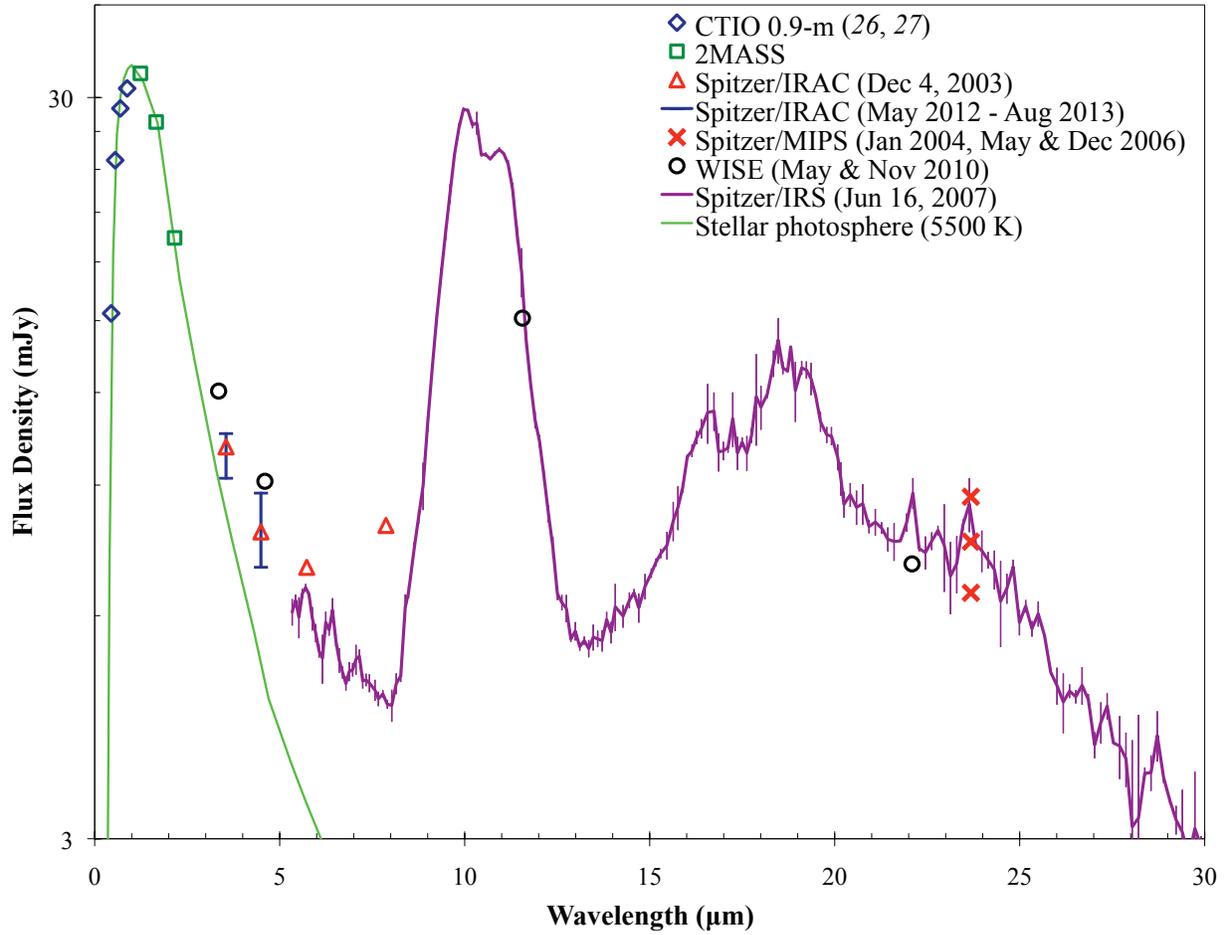

**Fig. 1**. Spectrum of the ID8 system. The photometric errors are smaller than the symbol size and are not shown; errors of the Spitzer/IRS spectrum are indicated as vertical lines. Red crosses at 24 μm show the three epochs of Spitzer/MIPS observations that led to the discovery of disk variability (*8*). The ranges of the new measurements at 3.6 and 4.5 μm in this work are plotted as blue vertical lines.

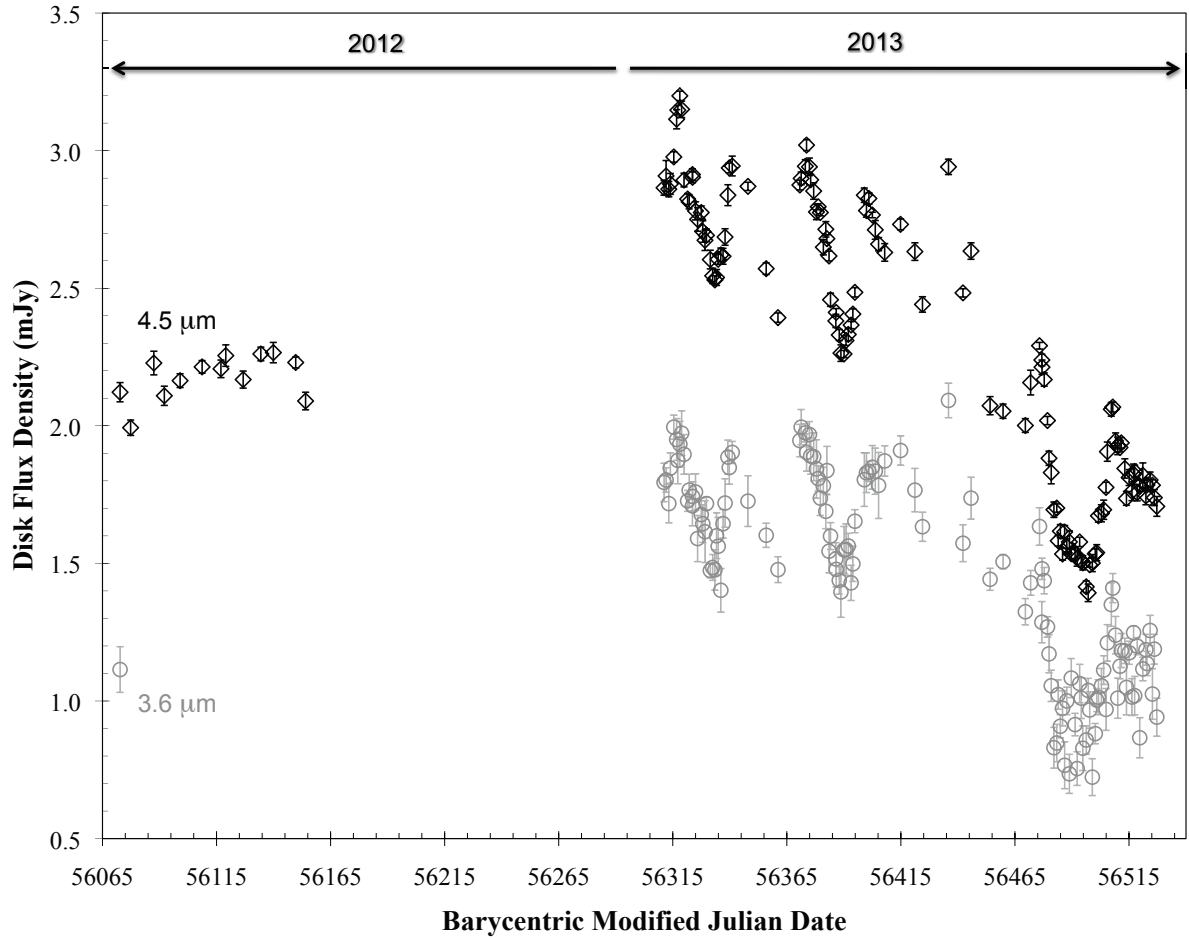

**Fig. 2**. Light curves of the debris disk of ID8 at 3.6 and 4.5 μm, assuming no error in photosphere subtraction. The gap between Barycentric Modified Julian Date (BMJD) 56155 and 56311 occurred when ID8 was outside of the Spitzer visibility windows.

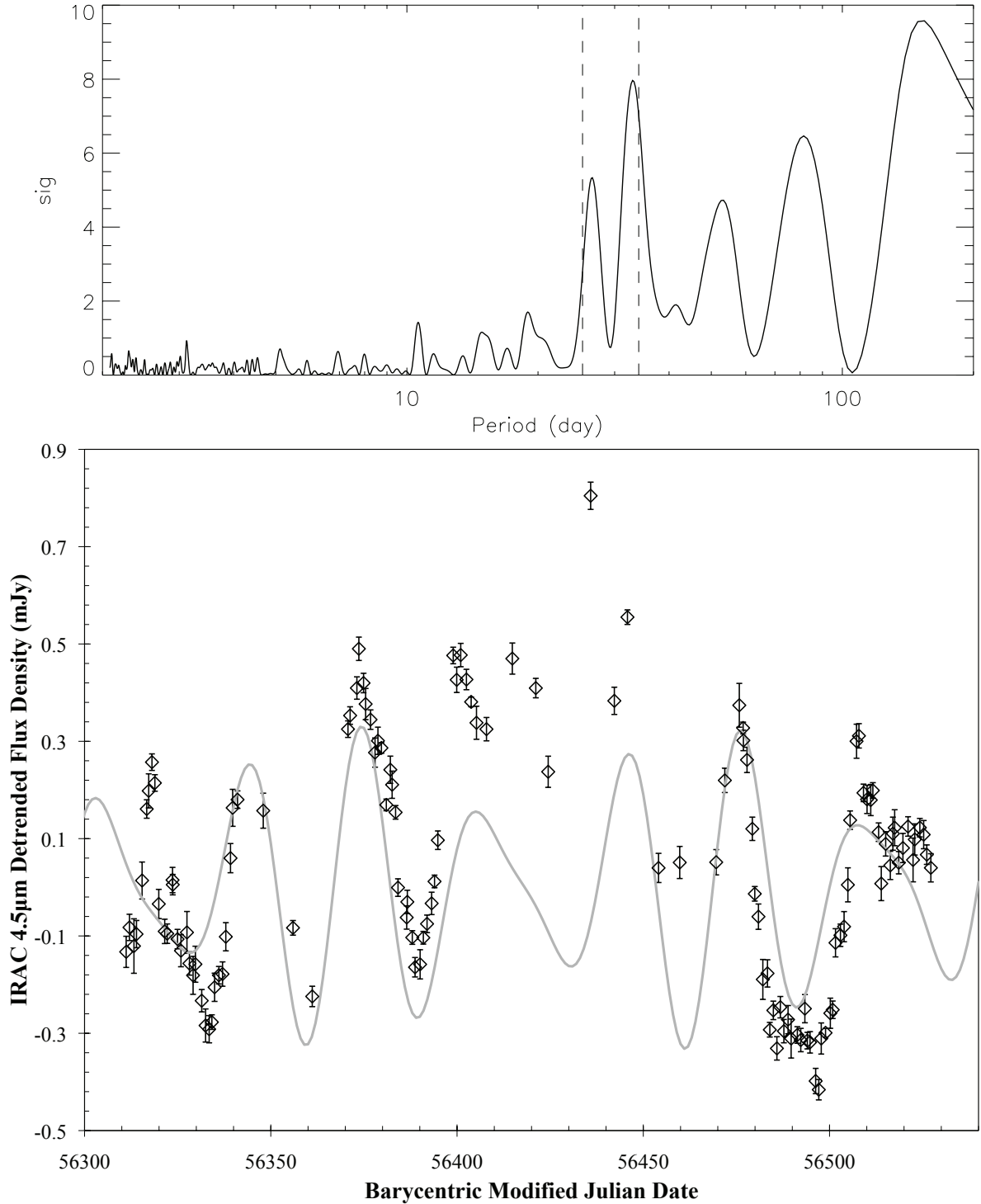

**Fig. 3**. Analysis of the 4.5 μm time-series of the ID8 disk. (**A**) Periodogram of the detrended 2013 light curve. The vertical dashed lines represent the identified periods $P_1$ and $P_2$, which are determined after removing the strongest frequency in each iteration (*22*) and thus appear slightly offset from the raw peak positions. (**B**) Detrended 4.5 μm data (black diamonds) with the composite of two sine waves of $P_1$ and $P_2$ (gray curve). Since their real waveforms are likely non-sinusoidal, the comparison is meaningful only for the general timing of the highs and lows.

# Supplementary Materials for

# Large Impacts around a Solar Analog Star in the Era of Terrestrial Planet Formation

Huan Y. A. Meng, Kate Y. L. Su, George H. Rieke, David J. Stevenson, Peter Plavchan, Wiphu Rujopakarn, Carey M. Lisse, Saran Poshyachinda, Daniel E. Reichart

correspondence to:  hyameng@lpl.arizona.edu

**This PDF file includes:**

Supplementary Text
Figs. S1 to S4

**Supplementary Text**

Optical Observations of ID8

Optical monitoring of ID8 was conducted in V and $I_C$ bands with the 0.41-m PROMPT 5 robotic telescope at Cerro Tololo Inter-American Observatory in Chile, from December 13, 2012 (BMJD 56292) to August 27, 2013 (BMJD 56531), completely covering the Spitzer visibility window in 2013. The typical cadence was 2-6 observations in each band throughout the night every night if conditions permitted, except for the period between July 16 and August 4 when ID8 was too close to the sun. All images were taken with an identical exposure time of 30 seconds. The CCD had a scale 0.60" pixel$^{-1}$ and a field-of-view of 10'. The pointing repeatability of the telescope was not perfect, making ID8 lie on different regions of the CCD every time, virtually equivalent to random dithering. Scientific images were prepared by an automatic pipeline with bias, dark, and flat field corrections, on which aperture photometry was performed with a radius of 5 pixels and sky annulus between 15 and 30 pixels. Two bright, unsaturated, and isolated stars near ID8, 2MASS J08085526-4856479 and 2MASS J08090978-4901041, were selected as reference stars for relative calibration. Neither star is a member of NGC 2547 (*28*). The former has a similar color as ID8 in the optical and near-infrared and was used as the calibration reference, while the latter is much redder and was used for verification purposes. A comparison of the two reference stars showed that both were stable within the ~1% measurement uncertainties over our entire time coverage.

ID8 appeared to be constant throughout our observations. The RMS scatters are 1.6% and 1.4% in V and $I_C$ bands, respectively. However, Fourier analysis reveals a significant period of 5.078±0.002 days with amplitude of 0.010 magnitude in both wavebands (Fig. S1). The shape, amplitude, and period are all consistent with the rotational behavior of other member stars with similar masses in the cluster (*29, 30*). In addition, given the distance of the cluster, spectroscopic and photometric observations both suggest that ID8 is a single star with no companion of similar mass. Therefore, we conclude that the stellar photosphere does not drive the disk variations, and the tiny optical variability is likely due to spots on the stellar surface brought into view by stellar rotation. The detection of a rotational period in the optical further suggests that the star is unlikely to be pole-on from our line of sight.

Spitzer Observations and Data Analysis

ID8 has a continuous ~221-day visibility window every year for the Spitzer Space Telescope, determined by the spacecraft pointing constraints. Spitzer/IRAC observations were made under programs 80260, 90192, and 90262, from May 25, 2012 to the end of the second visibility window on August 23, 2013, providing a total time baseline of 454 days with a 157-day gap. In 2012, observations were made at 4.5 μm only (one 3.6 μm observation was also taken), while all observations in 2013 used both wavebands. The basic sampling frequency was 1 observation every 7 days; three high cadence periods were obtained in 2013 with 1 observation every day for a total of 111 days.

We used a frame time of 30 seconds with 10 cycling dithering positions for both the 3.6 and 4.5 μm wavebands, achieving a typical signal-to-noise ratio of 150-300. The dithering pattern was designed to use a number of random pixel positions to average the intrapixel sensitivity variations of the detector (*31*). These data were processed with IRAC pipeline S19.1.0 by the Spitzer Science Center (SSC). We performed aperture photometry and compared the results using both the BCD (basic calibrated data) and mosaic post-BCD products. The BCD images have a native scale of 1.22" pixel$^{-1}$. An aperture radius of 3 pixels and a sky annulus of 12-20 pixels were used with aperture correction factors of 1.112 and 1.113 for 3.6 and 4.5 µm, respectively. The BCD photometry was corrected for the pixel solid angle (i.e., distortion) effects based on the measured target positions using files provided by the SSC. We also discarded any photometry when the target was too close to the edge of the detector array, and obtained weighted average photometry for each of the astronomical observation requests (AORs) by rejecting the highest and lowest photometry points in the same AOR. The same procedures were also conducted on the post-BCD products for comparison. We noticed that a few photometry points from the post-BCD products were significantly fainter (up to 10%) than the values obtained from the individual BCD images. Inspection of those AORs found that one or two BCD frames have poor WCS (world coordinate system) information, likely causing misalignment in the post-BCD products. Thus, we adopted our final photometry based on the weighted average of the BCD images.

To evaluate the uncertainty and stability of our photometry, we selected 4 stars in the field of view as references and obtained their photometry. These reference stars have similar or fainter fluxes than ID8. The measured RMS for them was smaller than 1% at both wavebands, consistent with the expected repeatability of the instrument (*31*, *32*). In comparison, the measurements of ID8 varied peak-to-peak by 16% and 23% at 3.6 and 4.5 µm during the same period, well beyond the instrumental errors. The consistency of the observations at 3.6 and 4.5 µm, which were made sequentially and used different detector arrays within IRAC and are therefore independent, confirms the reality of the variations.

The nominal photometric errors for our measurements, including the star and the disk, at individual epochs are typically 0.3-0.6%. However, the SED fitting suggests that the stellar photosphere contributes about 85% and 70% of the total output at 3.6 and 4.5 µm, respectively. After subtracting such large fractions of the total, the relative errors of the disk flux are 2-5%, and could be >10% at 3.6 µm when the disk is faint. More seriously, any error of the adopted photospheric flux would introduce a systematic error to the disk flux and color, which can hardly be identified in later analysis. Thus, the absolute values of the color temperature of the disk may be biased, but the relative color variations should be more robust. We calculated the expected trend of the entire ID8 system in a color-magnitude diagram, assumed the photospheric contribution is constant, and compared with two possible types of disk variations: 1) changing the dust emitting area at a fixed temperature, and 2) changing the dust temperature with a fixed emitting area. Changing stellar extinction could also affect the color of a system, but this scenario can be ruled out in the case of ID8 by the stability of the stellar flux in the optical, which should be much more sensitive to dust extinction. As illustrated in Figure S2, the observations show a continuous behavior with no abrupt changes in either dust area or temperature, slightly favoring the variations due to a combination of changing temperature and area or in area alone, rather than purely in temperature.

Frequency Spectrum Analysis

The time-domain characteristics of the ID8 disk, as in Figure 3(A), resemble the 3.6 and 4.5 µm light curves of some YSOs with protoplanetary disks, in which a characteristic timescale is present on top of significant long-term variability with no strict periodicity. However, the light curve patterns of YSOs are temporary, capable of switching from one quasi-period to another, or from quasi-periodic to stochastic within seasons (*33*). The rapid changes in protoplanetary disks are facilitated by gas accretion and strong magnetic interactions (e.g., *34, 35*); neither mechanism is applicable to debris disks, such as the one discussed in this paper. ID8 is much older than even the oldest known host stars of protoplanetary disks, and its stellar spectrum shows no trace of accretion. Therefore, changes of the type seen in the YSOs are not expected.

The frequency spectrum of the ID8 disk sees higher noise at the long period end ("red noise"), due to long-term deviations from the assumed decay function compared to the 216-day time baseline in 2013. In the periodogram in Figure 3(A), the most significant peak is the broad one centered around ~145 days. Because it is over two thirds of the time baseline and broader than an ideal monochrome, this feature more likely reflects a long-term deviation from the exponential decay rather than a real period. Second to the 145-day component, those at $P_2=34.0\pm1.5$ days and $P_1=25.4\pm1.1$ days come as the two most significant periods. Detailed modeling of sinusoidal functions with our sampling pattern showed that the broadening of their peaks in the periodogram is consistent with strict periodicity. (See the comparison in the upper panel of Fig. S3.) But the accurate periods depend on the assumed exponential detrending of the unweighted data, and are subject to change if different detrending and weighting are used. This is an additional source of errors that is not considered in the quoted errors, which should, therefore, be the lower limit of the real uncertainties of the periods. Thereafter, the next two orders would be periods of 18.7 and 13.3 days, which are visible in the lower panel of Figure S3 and might be overtones of $P_2$ and $P_1$. However, there is not enough information in the light curve to identify additional periods confidently.

There appear to be hints for 50- and 80-day periods in the original periodogram (Fig. 3(A)). However, both features disappear after the first three components (145-day, $P_2$, and $P_1$) are removed, as seen by comparing the panels in Figure S3. As a sanity check, using the exact sampling pattern obtained for the real observations, we found that the original frequency spectrum, including the 50- and 80-day features, can be reproduced simply by the first three components (Fig. S3(A)). This strongly suggests that the apparent additional peaks are artifacts caused by a combination of the stronger components and the specific sampling pattern (cf. *36*).

Frequency Spectrum Interpretation

We turned to the physical and dynamical model to address the characteristics of the ID8 light curve. The SED models suggest that the dust particles responsible for the ID8 disk emission are dominated by grains no larger than ~1 μm at about 0.4 AU from the star (*10*), where the dynamical timescale is consistent with a 71-day genuine period with $P_2$ and $P_1$ as the first and second order overtones. However, the evolution of the real disk involves various stochastic processes, like minor collisions that will cause random deviations from smooth variations of the disk flux. If there are other massive planets or planet embryos around, their perturbations will also affect the behavior of the disk. Translating these dynamical expectations into Fourier space, in the detrended light curve with the assumed exponential decay, one should expect to see an elevated floor of "red noise" and significant deviations like the 145-day component. These expectations are generally consistent with the observations.

To interpret the frequency spectrum of the disk flux, we need to consider the whole physical and dynamical picture. A critical constraint to any ID8 disk model is the short periods (~30 days) and high amplitude (~$6\times10^{-3}$) of the fractional luminosity variations described in the main text. If the dust detected in ID8 is optically thin, the quasi-periodicity of the light curve would suggest regular increases and decreases in the amount of dust. The decreases could be explained by radiation blowout, if a cloud of purely small grains gets ejected into the circumstellar region. However, detailed examination of mechanisms (e.g., repetitive multiple asteroid impacts, comets, evaporating planets, or giant volcanic eruptions) that produce dust cyclically shows them to be unlikely to be able to operate on a monthly basis and to produce the necessary large (~110 Jupiter radii) clouds. The timescale for variations in collisional cascades, where the small particles are produced by grinding down large bodies, are controlled by the evolution of the large bodies, which is much too slow to account for the behavior in ID8.

The disk mass estimate along with the grain size required to fit the mid-infrared spectrum indicates that the dust cloud should be optically thick, in which case variations are easier to understand. One simple explanation is the line-of-sight effect of an optically thick dust concentration, introduced in the main text. Here, we test whether this proposed scenario could produce the observed periodicity in the disk, characterized by strong first and second order overtones ($P_2$ and $P_1$) but a moderate or weak fundamental. Since there are too many unknowns in the ID8 planetary system, we did not attempt to make a perfect fit to the observations, but focused on the qualitative properties instead. Given the minimum disk mass and the average particle size of ~1 μm, the observed ID8 disk should contain ~$10^{33}$ dust particles. Computing for such a tremendous number of particles would be prohibitively expensive in time. To reduce the computing load, we used 5000 massless test particles and assumed a radius of 0.002 AU for each of them to make up the same total dust surface area as in the real disk of ~1 μm particles. Generally, we found that a generous range of mass, orbital, and dust expansion configurations is able to produce the desired qualitative results. The overtones tend to be more prominent for impacts occurring at farther distances and on moderately eccentric orbits. However, the azimuthal orientation of the line of sight makes little difference to the time domain characteristics of the disk flux.

As an example, here we show the simulation of an impact with a target of 1 Earth mass, placed on an orbit with semimajor axis a=0.327 AU, corresponding to the proposed 71-day genuine period. The orbital eccentricity is largely a free parameter, here assumed as e=0.6 to be compatible with the color trend that suggests a combination of varying temperature and varying dust amount as the most plausible cause of the ID8 disk variations. For the initial conditions of the dust, to simulate the outcome of an impact we placed all test particles at 50 Earth radii around the centroid of the planetary body when it is at aphelion, which means that the simulated single impact will be responsible for the entire disk flux. This moment is considered day 0; the time in which the dust particles escape to 50 Earth radii is neglected. The initial relative velocities of the test particles with respect to the impact target were assumed to follow a truncated Gaussian distribution with mean 3.23 km s$^{-1}$, standard deviation 3.81 km s$^{-1}$, and truncation threshold 1.67 km s$^{-1}$ (*37*, *38*). The impact ejection is likely anisotropic, but since we had no information on the preferred orientation, we opted to simulate isotropic ejection. Mutual collisions between dust particles were neglected. A dust particle was considered lost if it went out beyond 1 AU, but this rarely happened as we only focused on the first 500 days.

Simulations were carried out with the hybrid algorithm provided in the Mercury package (*39*), in which particles are integrated symplectically except for close encounters defined within 3 Hill radii from a massive body, when a Bulirsch-Stoer integrator takes over. The algorithm includes neither terms for tidal circularization from the host star, nor general relativity perturbations. However, at this separation and simulation timescale, these effects are negligible. The spatial position of each particle was computed for each day, and the total visible dust area was computed by projecting all positions to the celestial sphere according to the line of sight. The modeled disk flux density was in arbitrary units, set for convenience to the total visible dust area in AU$^2$ multiplied by a factor of 100. For simplicity, the line of sight was selected to be along the orbital major axis of the impact target at inclinations of 0° (exactly edge-on) and 30° above the orbital plane. The inclination to the orbital plane cannot be too large in order to produce the geometric effect.

The modeled light curves were purely geometric, ignoring the details of optical depth and radiative transfer, as well as the overall loss of particles and decay of the emission. As shown in Figure S4(A), the simulated light curves show large amplitude fluctuations shortly after the impact as the particle cloud is expanding. As we did not consider the loss of particles, the simulated light curves have a prolonged rising phase and do not decay as in the observations. However, the segment with stable flux after the rising phase should be equivalent to the detrended time-series in 2013. Therefore, we analyzed the simulated light curves between day 301 and 500 when the disk fluxes and variation patterns have been generally stable. The periodograms of these simulated light curves (Fig. S4(B)) reveal the 71-day dynamical period as a broad component, while the first and second order overtones are both sharp and significant, consistent with the identified periods $P_1$ and $P_2$ within the errors. A prediction of this model is that the harmonics will get weaker as the dynamical evolution of the dust cloud continues. By 1000 days after the impact, the fundamental period may appear stronger than the overtones. Ultimately, all periodic signals will vanish when the density fluctuations within the disk are smeared out.

The real ID8 disk is probably more complicated. For example, observations in the past decade may suggest a baseline of disk flux, which was not considered in our simulations. In addition, the initial particle sizes after the impact might be dominated by large condensates whose surface areas are smaller and thermal equilibrium temperatures are lower than those of the ~1 μm grains later generated by collisions. Such particles would radiate less efficiently per unit mass, making their output small in the first few orbits after the impact. The smaller particle sizes are close to our working wavelengths, where Mie theory suggests that the scattering cross section of a particle is greater than its geometric counterpart by a factor of ~2. Both possibilities would tend to suppress the output from the dust cloud when it first forms, compared with our geometric simulations. That is, it is possible that the impact occurred during or shortly before the 2012 observations but the emission from the resulting dust cloud was sufficiently weak that we did not detect it. Nonetheless, the periodicity in the simulation results is not qualitatively affected by these issues, and validates the line-of-sight effect as a viable explanation for the large amplitude of the sky coverage by dust.

Scale of the Impact

Now we can use all the available information in an attempt to understand the scale of the observed impact around ID8.

Near the end stage of terrestrial planet formation, rocky building blocks of a range of sizes are expected in the terrestrial planet region, ranging in size downward from a small number of large planetary embryos with lunar to Mars masses to objects with individual masses negligible compared to the embryos. "Giant impacts" refer to the collisions between two planetary embryos, which are expected to be outnumbered by planetesimal-embryo impacts and planetesimal-planetesimal collisions. Numerical simulations suggest that ~10 giant impacts are required for the formation of an Earth-like planet, while the Moon-forming impact is probably the last giant impact onto the Earth (*40*).

However, the impacts in the ID8 system are not necessarily a good analog of the Moon-forming event in the early solar system. The primary distinction with the canonical Moon formation model is that the Moon-forming impact is largely "closed," in which the impact fragments are mostly confined within the Earth's gravitational regime (*5*). Even if the entire Hill sphere of the proto-Earth were filled with optically thick dust after the impact, the infrared fractional luminosity of the dust would be under $10^{-5}$, i.e., 3 orders of magnitude lower than what we see in ID8. The Moon-forming event falls into the "graze-and-merge" category in impact models (*41*), which generally produces less debris compared to other types of impact outcomes, unless the combined body exceeded its spin stability limit (*7*). By contrast, given the high fractional luminosity of ID8, a large amount of debris must be spread into the interplanetary space. This might match the Moon-forming impact only if it was onto a fast-spinning proto-Earth, in which case a few to several percent of Earth mass can become unbound debris (*7*).

The nature of the ID8 impact can be estimated by comparing the frequency of similar events. Since its "extreme debris disk" was first observed in December 2003 (*28*), an impact must have occurred before then. In this work, we report a new impact before 2013. There might also have been an impact before May 2006 that caused the increase of the 24 μm flux of the disk (*8*). Thus, we can take 10 years as an approximate upper limit to the time between collisions. If this frequency arises from independent events, one would expect to find a high incidence of similar warm debris disks around other solar-like stars with ages in the era of terrestrial planet formation. However, an unbiased Spitzer survey of young open clusters in the age range of 30-130 Myr found "extreme debris disks" around only 1% of solar-like stars (*42*). A similar result was also found in a census of nearby solar-like stars in this age range (*43*). The Kepler mission found that >17% of solar-like stars have at least one Earth-mass planet (*44*), so avoiding this contradiction by assuming ID8 is a rare example of planetary systems is not viable.

Another approach to estimate the nature of the ID8 impact is by the duration of the impact consequences. We can translate the 1% disk incidence (*42*, *43*) to an estimate of impact durations. To order of magnitude, during the ~100-Myr-long era of terrestrial planet formation, the average solar-like star should spend a total of 1 Myr hosting "extreme debris disks". If the time between independent ID8 impacts is <10 years, there will be $>10^7$ collisions over the age range. To make up the 1 Myr total observable time, the average duration of the observable phase after each collision should be <0.1 year. This clearly contradicts the observations. All known "extreme debris disks" have been sustained for one to a few decades; even the only known disappearing disk had a lifetime of at least 27 years before it vanished (*45*). (We just have the infrared technology to observe them for a few decades. The actual duration of these disks could be orders of magnitude longer.) Therefore, the ID8 impacts before 2003 and before 2013 cannot be independent. They are probably related events triggered by an original impact, which placed debris on orbits that enhance the incidence of secondary collisions. Such secondary collisions may be (a) re-accretion events of previously escaped fragments from the original impact, (b) collisions between two escaped fragments, or (c) collisions of escaped fragments with other planetesimals. Among them, (a) and (b) are expected events in the aftermath of an original impact based on dynamical arguments (*38*), while the incidence of (c) depends on the spatial density of ambient planetesimals.

Though it is clear that the 2013 impact around ID8 is secondary, the scale of the original impact is poorly constrained by the current data. The originator of the ID8 excess could be an embryo-embryo "giant impact," about ten of which are expected to make an Earth-like planet. However, an original impact of smaller scale is equally consistent with the available constraints. Consider a planetesimal swarm of N bodies each of radius R km. To order of magnitude, the total mass is $m \sim N R^3/10^{11.5}$ in Earth masses. In the presence of larger planetary embryos that excite relative velocities of 10-20 km s$^{-1}$, the time between mutual planetesimal collisions is $\tau \sim 10^{16}/(R^2 N^2)$ in years. If we assume that the original impact is between two large planetesimals of radius 500 km, the observable phase with dust-replenishing secondary collisions after the original impact may be 1000 years, including the one we see before 2013. To make the 1 Myr total observable time, we will need 1000 primary impacts over the era of terrestrial planet formation. This requires about 6000 such planetesimals with a total of 2.5 Earth masses. These numbers are also consistent with our observations and reasonable within our understanding of terrestrial planet formation.

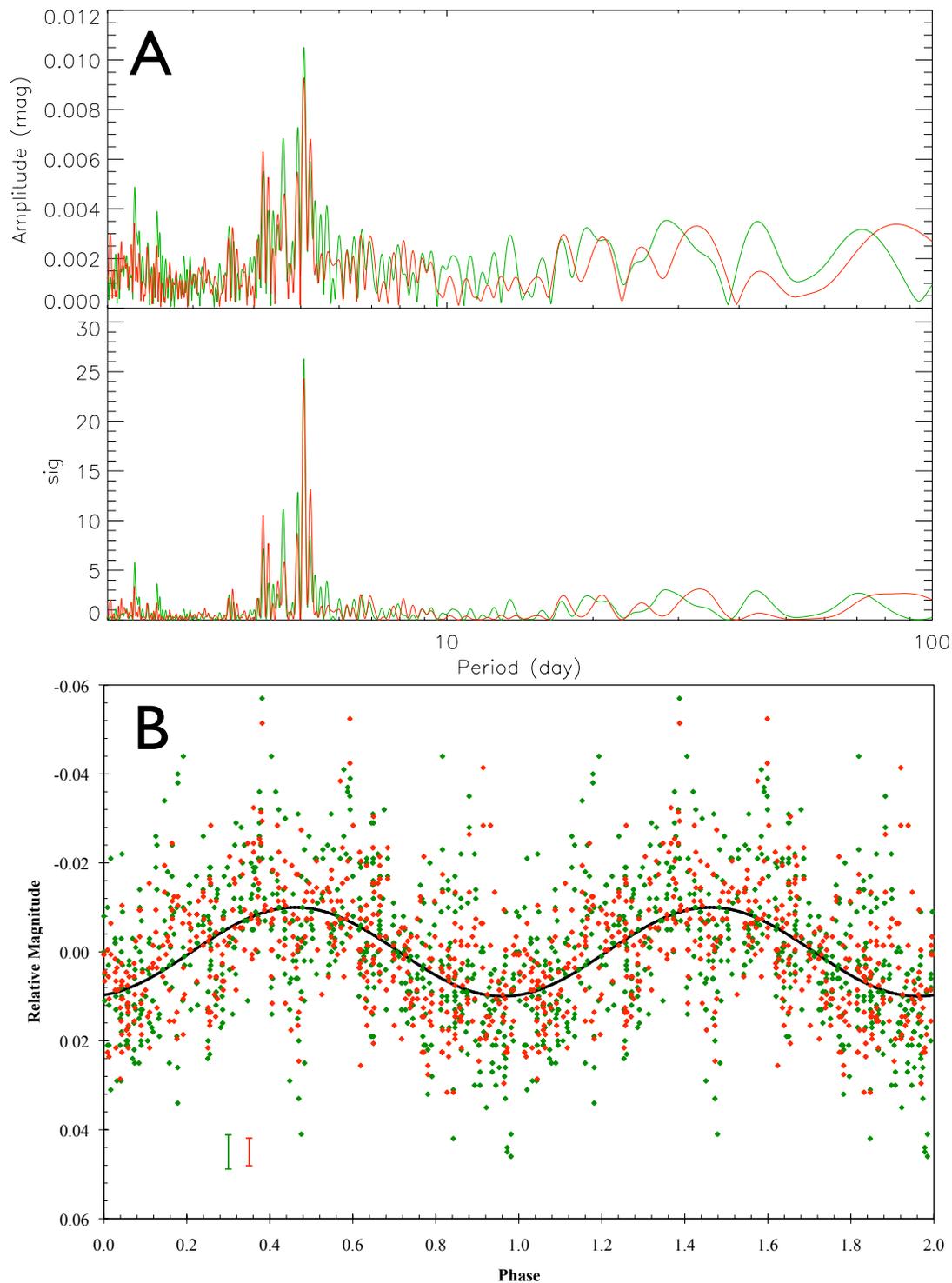

**Fig. S1.**

Analysis of the optical light curve of ID8 in V (green) and $I_C$ bands (red). (**A**) Frequency spectrum. (**B**) Phase-folded curve with the 5.078-day period. Nominal photometric errors are plotted on the lower left corner. The solid black line is a sine wave with the same phase and amplitude for comparison.

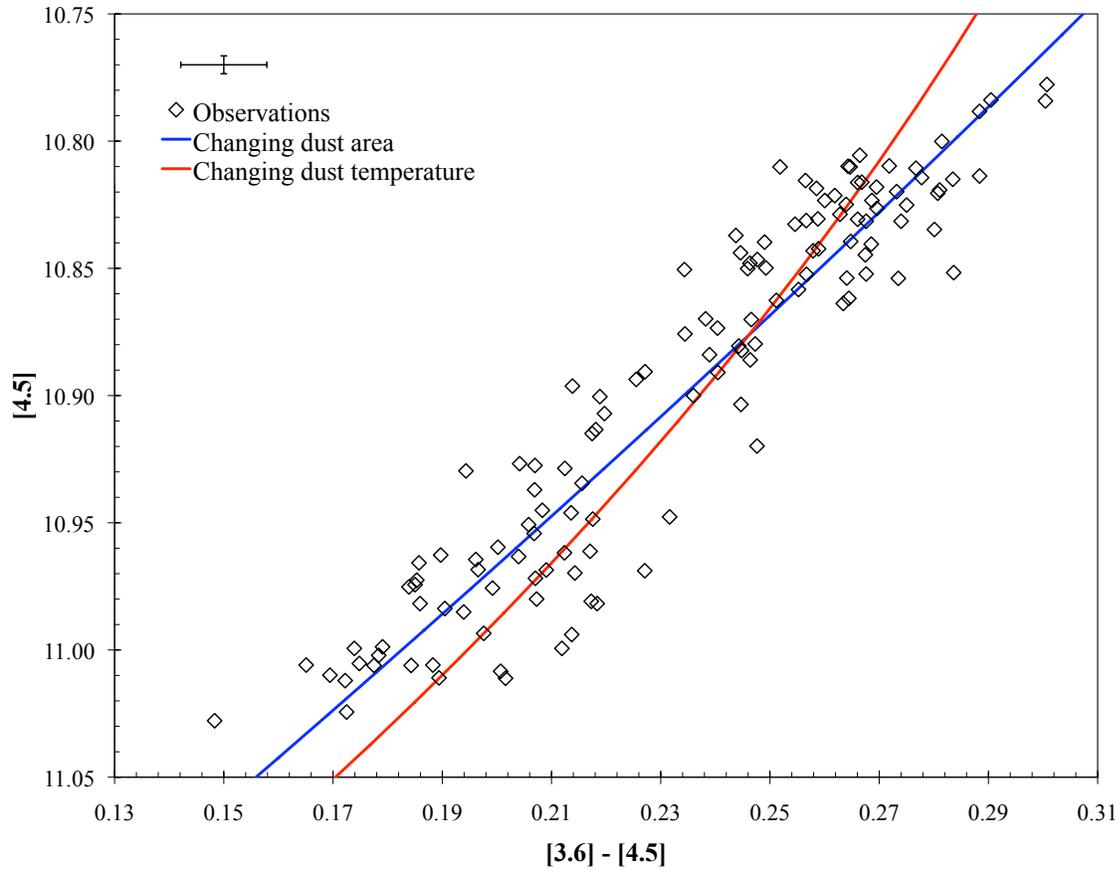

**Fig. S2**

Color-magnitude diagram of the ID8 debris disk. The average error is in the upper left corner. The Y-axis is the brightness relative to Vega in astronomical magnitudes, while the X-axis is the difference in magnitudes at 3.6 and 4.5 μm. Assuming a constant stellar photosphere and blackbody for the dust, theoretical tracks are overplotted by changing the dust emitting area at a fixed temperature, and changing the dust temperature with a fixed dust emitting area.

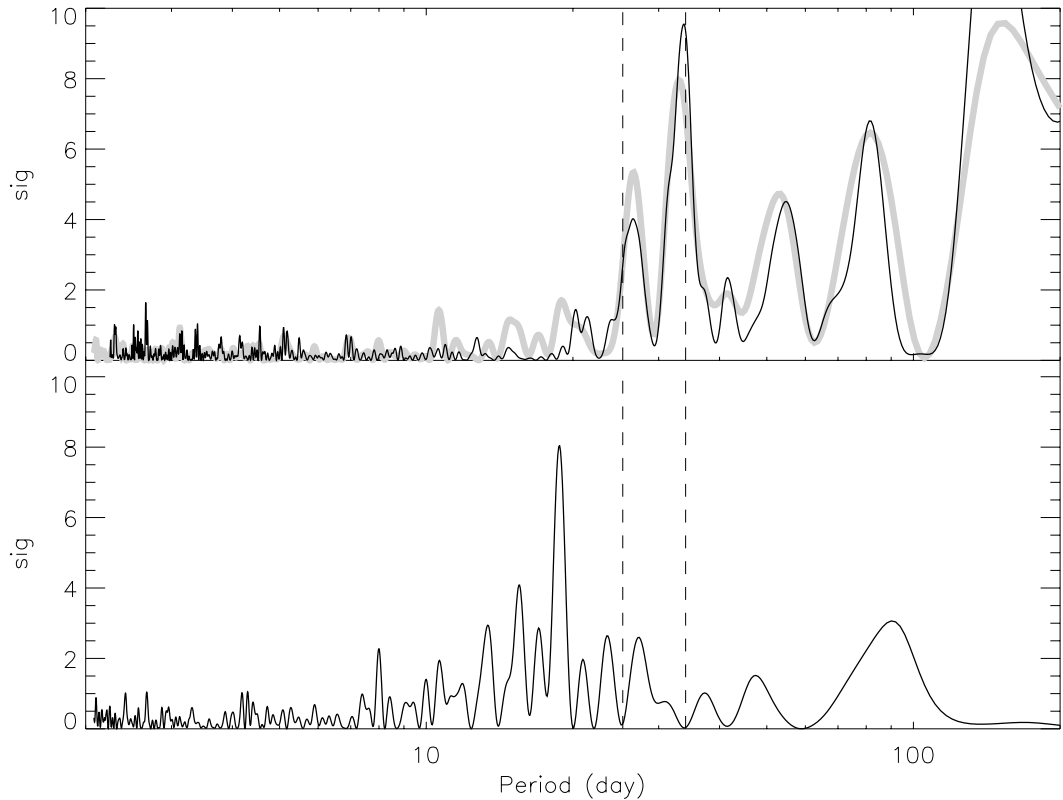

**Fig. S3**

Frequency spectrum analysis. Upper: Periodogram of the simulated light curve for the detrended 4.5 μm times-series in 2013 (black line). The light curve is reproduced by a combination of three sine waves with the observed amplitudes and phases of the 145-day component, $P_2$, and $P_1$, plus normally distributed photometric noises at the same level as in the observations. The vertical dashed lines represent the positions of $P_1$ and $P_2$. For comparison, the spectrum of the detrended real observations (a replica of Fig. 3(A)) is overplotted as a thick gray line. The 50- and 80-day features, though not parts of the simulation input, are reproduced comparably to those in the real observations. Lower: Residual spectrum of the real observations after the 145-day, $P_1$, and $P_2$ are removed as sine waves. The 50- and 80-day artifacts are also gone.

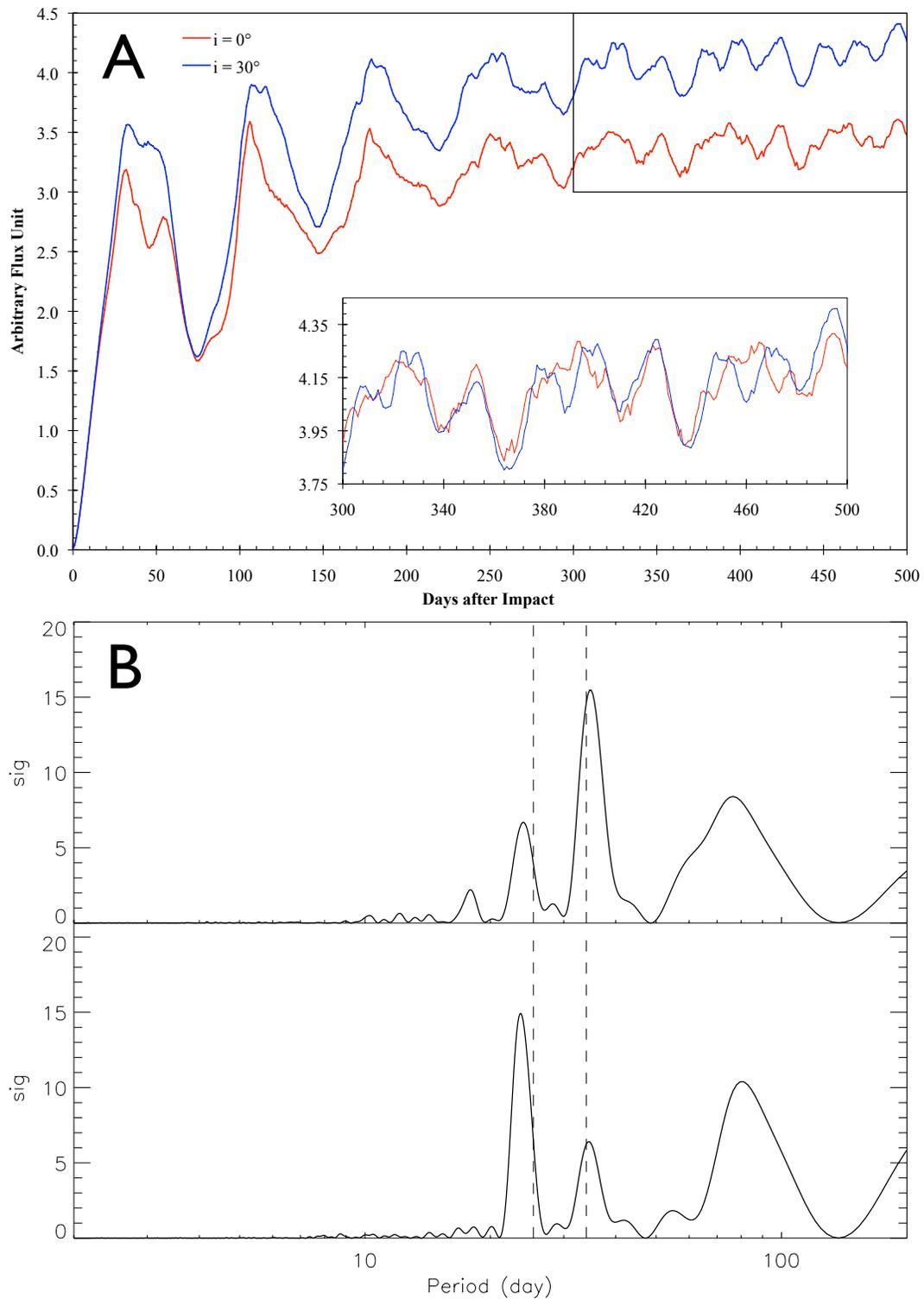

**Fig. S4**

Simulations for the line-of-sight effect. **(A)** Simulated light curves. The red line is for the view along the orbital major axis through the orbital plane (0°) and the blue line is viewed from an inclination of 30°. Segments embraced by black rectangles are magnified in the embedded chart with the 0° curve shifted +0.709 flux unit for a better comparison, and are the region used for the periodograms. **(B)** Periodograms of the simulated light curves. The upper and lower panels are for 0° and 30° inclinations, respectively. The vertical dashed lines mark the positions of $P_1$ and $P_2$, which are not input to the simulation.